# Mass Prediction of the Weak and Higgs Bosons Using the Massless Spin-Exchange Preons Model: An Alternative to the Higgs Mechanism in Yang-Mills Theory

Jau Tang[1,*] and Qiang Tang[2]

[1]Institute of Technological Sciences, Wuhan University, Wuhan 430074, China

[2]Anhui University of Science and Technology, Huainan, Anhui 232000, China

[*]Corresponding author: Jau Tang, WuhanTang72@gmail.com

**Abstract**

We modify the Yang-Mills model using massless preon pairs to describe the photon isospin singlet and the Z and W bosons. Unlike the Higgs mechanism, which depends on a scalar field our model employs Gell-Mann generators, revealing that the masses of W and Z bosons arise from the internal dynamics of chiral preon pairs. With no adjustable parameters, we predict $m_W/m_Z = \sqrt{3/2} \approx 0.87$ (experimental value: 0.88), a Weinberg angle of 30° (vs. 29°), and a decay width ratio of $\Gamma_W/\Gamma_Z = \sqrt{3/2} \approx 0.87$ (vs. 0.84). Additionally, we find $m_H/m_W = \sqrt{7/3} \approx 1.53$ (vs. 1.56) for the Higgs boson. Our approach introduces a curl operator leading to chirality, characterized by a topological structure of Möbius tori, providing new insights into particle interactions and challenging existing theoretical paradigms.

In the Standard Model[1-3] of particle physics, the W and Z vector bosons[4-5], as the carriers for the weak force, are considered point-like elementary particles and are not composed of smaller constituents. However, such a line of thinking could not explain the origins of the three generations[6] of leptons and quarks[7], their mass ratios which follow a simple Koide formula.[8] It is unclear why neutrinos have a small mass[9] and why some particles remain massless even though the Higgs field[10-11] is omnipresent. Different from the Yang-Mills theory[12] in the conventional electroweak theory,[13] we propose a model based on massless chiral preons as a building block for the isospin singlet and triplet vector bosons. According to this model, an interacting massless isospin-1/2 preon pair could form a massless singlet with oppositely aligned spins, and a massive weak boson triplet with parallel spins. The singlet-triplet energy splitting is a result of spin-spin exchange interaction. Therefore, the photon possesses spin-1 and isospin-0, but the weak bosons own spin-1 but isospin-1.

Our preon-pair idea originates from the singlet-triplet formation and splitting in molecular physics.[14,15] To explain the formation of a photon as an isospin singlet and a weak boson as a triplet in a family, we propose this family of four is formed by two doublets so that $2 \times 2 = 3 \oplus 1$, like a singlet-triplet family in quantum theory for chemical bonds, except that the spin is an isospin related to internal degrees of freedom. For two spin-1/2 vectors, $s_1$ and $s_2$. The v0, depending $S = s_1 + s_2$, and one has $S^2 \equiv S \cdot S = s_1 \cdot s_1 + 2s_1 \cdot s_2 + s_2 \cdot s_2$, which leads to $S^2 = S(S+1) = 2$ or $0$, depending on whether two spins are parallel with $S = 1$ or opposite with $S = 0$, where $s_1 \cdot s_1 = s_2 \cdot s_2 = 3/4$ for spin-1/2 particles. For the triplet, there ae three states, $|\uparrow\uparrow\rangle, (|\uparrow\downarrow\rangle + |\downarrow\uparrow\rangle)/\sqrt{2}, |\downarrow\downarrow\rangle$, and for the singlet there is one state $(|\uparrow\downarrow\rangle - |\downarrow\uparrow\rangle)/\sqrt{2}$. The energy splitting between the triplet and the singlet is caused by the exchange interaction involving $s_1 \cdot s_2$ between two spin-1/2 particles with the eigenenergy given by $E = \Delta(1/2 + 2s_1 \cdot s_2)/2$ so that the singlet with $S = 0$ is at zero energy and the triple with $S = 1$ is at energy at $\Delta$.

We shall show that both W and Z bosons possess an internal structure and acquire their rest masses via the exchange interactions between the paired preons. Unlike the Standard Model, our views of these vector bosons as composites are supported by the existence of the photon as an isospin singlet, the weak boson as a triplet, and the Koide mass ratio for charged leptons and quarks. Our dual-preon model for the vector bosons offers an alternative to the conventional electroweak theory which treats them as a point-like object, it also differs from the Higgs mechanism that invokes a spontaneously broken scalar field. We shall show that the interacting preo pairs with opposite chirality have an internal structure with their mass acquired from internal kinetic energy. We shall demonstrate the advantage of this by making fairly accurate prediction, with no adjustable parameters, of their mass ratios among W, Z, Higgs bosons, and the decay width of the weak bosons.

For a massless particle according to our dual-component model, we consider a wave function $\Psi(t,x)$ composed of a pair of real-value functions $f(t,\mathbf{r})$ and $g(t,\mathbf{r})$ governed by the following wave equation

$$\left(-\frac{\partial^2}{\partial t^2}^2 + \frac{\partial^2}{\partial x^2} + \frac{\partial^2}{\partial y^2} + \frac{\partial^2}{\partial z^2}\right)\Psi(t,\mathbf{r}) = 0, \Psi(t,t) \equiv \begin{pmatrix} f(t,\mathbf{r}) \\ g(t,\mathbf{r}) \end{pmatrix} \quad (1)$$

By Fourier transformation of the above wave equation one can obtain Einstein's relation for a photon with $E = \hbar\omega, p = \hbar\omega, \omega = ck.$ In a later discussion, $\hbar = c = 1$ will be used. The above equation is a special case of the Klein-Gordon equation for a scalar boson without a rest mass.

In this work, we shall present a dual-component model for the unified treatment of electroweak interactions. According to this model, there are two types of equations, one for the boson-type preon and the other for the fermion-type preon. A boson-type preon contains two degrees of freedom and can be represented by two real-valued wave functions. A fermion-type preon is a spinor constructed from a pair of boson-type preons. We first consider

$$\begin{array}{cc} \mathbf{f}(t,\mathbf{r}) & \mathbf{g}(t,\mathbf{r}) \end{array}$$
$$\frac{\partial}{\partial t}\begin{pmatrix} \mathbf{f}(t,\mathbf{r}) \\ \mathbf{g}(t,\mathbf{r}) \end{pmatrix} = -(\nabla \times)\begin{pmatrix} 0 & -1 \\ 1 & 0 \end{pmatrix}\begin{pmatrix} \mathbf{f}(t,\mathbf{r}) \\ \mathbf{g}(t,\mathbf{r}) \end{pmatrix} \quad (2A)$$
$$\mathbf{f}(t,\mathbf{r}) = \begin{pmatrix} f_1(t,\mathbf{r}) \\ f_2(t,\mathbf{r}) \\ f_3(t,\mathbf{r}) \end{pmatrix}, \mathbf{g}(t,\mathbf{r}) = \begin{pmatrix} g_1(t,\mathbf{r}) \\ g_2(t,\mathbf{r}) \\ g_3(t,\mathbf{r}) \end{pmatrix}$$
$$\sigma_t \equiv \begin{pmatrix} 0 & -1 \\ 1 & 0 \end{pmatrix}, \sigma_t^2 \equiv -I_2$$

where $\varepsilon_{ijk}$ is the Levi-Civita symbol for the cross-product operation. The skew-symmetric matrix $\sigma_t$ plays a role like the imaginary number in the conventional quantum theory that uses complex wave functions. If one defines a complex wave function one can rewrite the above equation as

$$i\,\partial/\partial t\,\Psi(t,\mathbf{r}) = (\nabla \times)\Psi(t,\mathbf{r}), \Psi(t,\mathbf{r}) \equiv \mathbf{f}(t,\mathbf{r}) + i\mathbf{g}(t,\mathbf{r}) \quad (2B)$$

Eqs. (2A-2B) represents the 1$^{st}$-order differential wave equation for a massless "preon" which has dual-component real-value wave functions. Taking its 2$^{nd}$-order time derivative, one obtains

$$\partial^2/\partial t^2\, \pmb{\Psi}(t,\pmb{r}) = -\nabla \times (\nabla \times \pmb{\Psi}(t,\pmb{r})) = -\nabla(\nabla \cdot \pmb{\Psi}(t,\pmb{r})) + \nabla^2 \pmb{\Psi}(t,\pmb{r}) = \nabla^2 \pmb{\Psi}(t,\pmb{r}) \quad (2C)$$

where $\nabla \cdot \pmb{\Psi}(t,\pmb{r}) = 0$, or $\nabla \cdot F(t,\pmb{r}) = \nabla \cdot G(t,\pmb{r}) = 0$ was assumed in vacuum-like the conditions for an electric and magnetic field. Eq. (2C) indicates that such a preon is massless. The curl operator in Eqs. (2A) - (2B) coupes cyclically each axial component to two other component. It has a topological structure like three strains of intertwined fiber bundles of Möbius tori, which is a 3D extension of a 2D Möbius strip.[16] This topological structure underlies the three-color concepts for quarks and gluons and is related to a branch of differential topology called the Hopf fibratio.[17]

In addition to the boson-type preon, we now consider the fermion-type preon. Similar to Dirac's operator approach[18,] except that a curl operator is used here, we incorporate four anti-commutative matrices into the momentum operators in Eq. (2B) to obtain

$$i\frac{\partial}{\partial t}\pmb{A}_0 \otimes \pmb{\Psi}_i(t,\pmb{r}) = \sum_{j,k=1}^{3} \varepsilon_{ijk} \frac{\partial}{\partial x_j} \pmb{A}_j \otimes \pmb{\Psi}_k(t,\pmb{r}), i,j,k = 1,2,3 \quad (3A)$$

where $\otimes$ represents a tensor product and the operator $\pmb{A}_\mu, \mu = 0,1,2,3$, are related to Dirac's gamma matrices which follow anti-commutative relations of $\{\pmb{A}_\mu, \pmb{A}_\nu\} = 2\delta_{\mu\nu}\pmb{I}_4, \mu, \nu = 0,1,2,3$. By taking the 2$^{nd}$-order time derivative of Eq. (3A) and the use of Eq (3B) we obtain

$$-\pmb{A}_0^2 \otimes \frac{\partial^2}{\partial t^2} \pmb{\Psi}_i(t,\pmb{r}) = \sum_{j,k}\sum_{l,m} \varepsilon_{ijk}\,\varepsilon_{klm} \backslash \pmb{A}_j \pmb{A}_l \otimes \partial^2 \pmb{\Psi}_m(t,\pmb{r})/\partial x_j \partial x_l \quad (3B)$$

Using the relations for summing up two Levi-Civita symbols and with the anti-commutative relations, one obtains $\pmb{I}_4 \otimes (\partial^2/\partial t^2 - \nabla^2)\pmb{\Psi}(t,\pmb{r}) = 0$, where $\nabla \cdot \pmb{\Psi}_i(t,\pmb{r})$ vanishes in vacuum. The 2$^{nd}$-order derivative wave equation for either boson- 0r fermion-type preon leads to Eq. (1) for a massless particle.

In a quantum system, two spin-1/2 particles can form a singlet and triplet, i.e., $\pmb{2 \times 2 = 3 \oplus 1}$, we use such a pair of massless spin-1/2 preons with an opposite chirality to construct the massless singlet photon and the massive weak boson triplet. In such a composite system with a preon at $\pmb{r}_1$ and the other at $\pmb{r}_2$, one can define the average position $\pmb{r} = (\pmb{r}_1 + \pmb{r}_2)/2$ and the relative position $R = (r_1 - r_2)/2$. For a singlet preon pair without an internal structure, one only needs to describe its dynamics. Let's consider the following equation for a pair of real-value paired wave functions $(f(t,r), g(t,r))$ and $(F(t,r), G(t,r))$

$$-\frac{\partial}{\partial t}\begin{pmatrix}\Psi_1(t,r)\\\Psi_2(t,r)\end{pmatrix} = (\nabla \times)\begin{pmatrix}\sigma_t & 0\\0 & -\sigma_t\end{pmatrix}$$
$$\Psi_1(t,r) = \begin{pmatrix}f(t,r)\\g(t,r)\end{pmatrix}, \Psi_1(t,r) = \begin{pmatrix}F(t,r)\\G(t,r)\end{pmatrix} \quad (4A)$$

Using complex-vale wave functions $\Psi_1(t,r) = f(t,r) + i\,g(t,r), \Psi_2(t,r) = F(t,r) + i\,G(t,r)$ the above equation becomes

$$i\frac{\partial}{\partial t}\Psi(t,\mathbf{r}) = (\sigma_3 \otimes I_2)\nabla \times \Psi(t,\mathbf{r}), \Psi(t,\mathbf{r}) \equiv \begin{pmatrix}\Psi_1(t,\mathbf{r})\\\Psi_2(t,\mathbf{r})\end{pmatrix} \quad (4B)$$

The above equation leads to $\nabla \times g = \partial f/\partial t$, $\nabla \times f = -\partial g/\partial t$, $\nabla \cdot f = 0, \nabla \cdot g = 0$ for the preon. For the preon with an opposite chirality one has $\nabla \times F = \partial G/\partial t$, $\nabla \times G = -\partial F/\partial t$, $\nabla \cdot F = 0, \nabla \cdot G = 0$. We can regard a photon as a singlet of an opposite-chirality preon pair with no internal structure. By assigning the electric field $\mathbf{E}(t,\mathbf{r})$ to $\Psi_1(t,\mathbf{r})$ and the magnetic field $\mathbf{B}(t,\mathbf{r})$ to $\Psi_2(t,\mathbf{r})$, Eq. (4B) is exactly Maxwell's equation in a vacuum without a source.

Unlike the photon which is an isospin-0 singlet with no internal degrees of freedom, , we construct an isospin triplet from an interacting massless preon pair with opposite chirality. We shall extend the dual-component model to the triplet weak vector bosons. In comparison to the photon singlet, the weak bosons are represented by a triplet. Therefore, we consider them having an internal degree of freedom. For a triplet with isospin-1, we employ Gell-Mann's 3x3 lambda matrices to describe the couplings in their internal dynamics, involving a coordinate vector $R$. To distinguish the two sets of coordinates, we denote $\nabla_r$ and $\nabla_R$ as the corresponding gradients. We consider the following equation for a pair of massless preon $\Psi_1(t,\mathbf{r},R)$ and a $\Psi_2(t,\mathbf{r},R)$ with an opposite chirality

$$i\frac{\partial}{\partial t}\Psi(t,\mathbf{r},R) = (\sigma_3 \otimes I_3)\nabla_r \times \Psi(t,\mathbf{r},R) + \left(\sum_{n=1}^{3} Q_n \sigma_1 \otimes \Lambda_n + Q\sigma_2 \otimes \Lambda_9\right)\nabla_R \times \Psi(t,\mathbf{r},R)$$
$$\Psi(t,\mathbf{r},R) = \begin{pmatrix}\Psi_1(t,\mathbf{r},R)\\\Psi_2(t,\mathbf{r},R)\end{pmatrix} \quad (5A)$$

where the operators $\Lambda_n$ are related to a subset of Gell-Mann's 3x3 SU(3) generator[1-2] matrices which are defined below[18]

$$\Lambda_1 = \begin{pmatrix} 0 & 1 & 0 \\ 1 & 0 & 0 \\ 0 & 0 & 0 \end{pmatrix}, \Lambda_2 = \begin{pmatrix} 0 & -i & 0 \\ i & 0 & 0 \\ 0 & 0 & 0 \end{pmatrix}, \Lambda_3 = \begin{pmatrix} 1 & 0 & 0 \\ 0 & -1 & 0 \\ 0 & 0 & 0 \end{pmatrix}, \Lambda_8 = \frac{1}{\sqrt{3}}\begin{pmatrix} 1 & 0 & 0 \\ 0 & 1 & 0 \\ 0 & 0 & -2 \end{pmatrix}$$

$$\Lambda_0 = \begin{pmatrix} 1 & 0 & 0 \\ 0 & 1 & 0 \\ 0 & 0 & 0 \end{pmatrix}, \Lambda_9 \equiv \frac{2}{\sqrt{3}}\begin{pmatrix} 0 & 0 & 0 \\ 0 & 0 & 0 \\ 0 & 0 & -1 \end{pmatrix} = \Lambda_8 - \sqrt{\sum_{k=1,2,3} \Lambda_k^2}$$

(5B)

where $[\Lambda_k, \Lambda_9] = \{\Lambda_k, \Lambda_9\} = 0, \{\Lambda_i, \Lambda_j\} = 2\delta_{ij}\Lambda_0, i,j = 1,2,3$. The set of three matrices $\Lambda_i, i = 1,,2,3$ belongs to SU(2) group, together with $\Lambda_0$ are a subset of SU(3) generators which were used in Gell-Mann's quark model. These three operators $\sigma_3 \otimes I_3$, and $\sigma_2 \otimes \Lambda_9$ anti-commute, and $Q_1 = Q \sin\theta \cos\phi, Q_2 = Q \sin\theta \sin\phi, Q_3 = Q \cos\theta$. By taking the 2$^{nd}$-order time derivative of Eq. (5A) and by Fourier transformation, one obtains

$$(\omega^2 - k^2)(I_2 \otimes I_3) = M^2, M^2 = Q^2 K^2 I_2 \otimes \begin{pmatrix} 1 & 0 & 0 \\ 0 & 1 & 0 \\ 0 & 0 & 4/3 \end{pmatrix}$$

(6)

where we have used $[\Lambda_k, \Lambda_9] = \{\Lambda_k, \Lambda_9\} = 0$ and an equality relation for the curl operator $-(\nabla \times)(\nabla \times)\Psi = -\nabla(\nabla \cdot \Psi) + \nabla^2\Psi = \nabla^2\Psi$ and the divergence $\nabla_r \cdot \Psi(t, r, R) = \nabla_R \cdot \Psi(t, r, R) = 0$ in vacuum. The coupling strength $Q$ for the singlet-triplet splitting is related to the spin-spin exchange interaction ting energy is between the isospin-1/2 preon pair with the energy $E = \Delta (1/2 + 2s_1 \cdot s_2)/2 = S(S+1)/2$ so that the singlet-triplet gap is $\Delta$.. The further splitting between the W boson doubles and the Z boson is due to the coupling involving $(1/2 - 2 s_{1z}s_{2z})$. In our model we employ $\Lambda_8$, the 8$^{th}$ Gell-Mann matrix, which is a diagonal matrix with a trace $tr(\Lambda_8^2) = 2$. This value corresponds to the square of the spin-1 angular momentum $S^2 = S(S+1) = S_x^2 + S_y^2 + S_z^2 = 2I_3$ where $S_y^2 = \Lambda_{8,22}^2 = 1/3$, $S_x^2 = \Lambda_{8,11}^2 = 1/3$, and $S_z^2 = \Lambda_{8,33}^2 = 4/3$.

According to Einstein's mass-energy relation, $E^2 - p^2 - m_0^2 = 0$, we obtain tan effective mass-squared diagonal matrix $M^2$. The above result implies that the exchange coupling between the paired isospin-1/2 preons has an anisotropic spin-spin coupling, often seen in magnetic materials, instead of Heisenberg's simplest isotropic coupling. Such an anisotropy breaks the triplet degeneracy with $M^2$ with $m^2 = Q^2 K^2 (2S(S+1) - S_z^2)/3 = Q^2 K^2 (2S \cdot S - S_z^2)/3$, which leads to $m = 0$ for the photon singlet $|0,0\rangle$, $m = QK$ for the $W^\pm$ bosons $|1, \pm 1\rangle$, and $m = QK2/\sqrt{3}$ for the Z boson $|1,0\rangle$. One can express the mass-square as $m^2 = Q^2 K^2 (1 + (4s_1 \cdot s_2 - (s_{1z} + s_{2z})^2)/3)$. showing the anisotropic exchange interaction between the spin-1/2 preon pair. Our model predicts a mass ratio $m_W/m_Z = \sqrt{3}/2 \sim 0.8660$, and a Weinberg angle $\theta_W$ of $30^0$. Given the rest mass for Z boson $m_Z = 91.1776$ Gev/c². For W boson $m_W = 80.377$ Gev/c²,[19] one has $m_W/m_Z = 0.8815$ and $\theta_W = 29^o$. According to the Higgs mechanism, these parameters require experimental measurements, yet they are derived theoretically from our model with no adjustable parameters..

We question the commonly accepted notion of the Higgs boson as an elementary God's particle because it is heavier than W, Z bosons, leptons, and some quarks. To show that it is a composite particle with $m_H = \sqrt{m_Z^2 + m_W^2}$, we consider the mixing between Z and W bosons in the following equation that involves only time dependence

$$i \partial \Psi_H / \partial t = (m_W \sigma_1 + m_Z \sigma_2) \Psi_H + \xi \sigma_3 \Psi_H, \quad \Psi_H = \begin{pmatrix} \Psi_W \\ \Psi_Z \end{pmatrix}$$
$$- \partial^2 \Psi_H / \partial t^2 = (m_W^2 + m_Z^2 + \xi^2) \Psi \tag{7}$$

where $\Psi_W = (\Psi_{W^+} - \Psi_{W^-})/\sqrt{2}$ is a coherent state of $W^\pm$. One obtains $m_H^2 = m_W^2 + m_Z^2$ if $\xi = 0$, indicating the Higgs boson is a composite. W $m_W/m_Z = \sqrt{3}/2$, we predict $m_H/m_W = \sqrt{7/3} \sim 1.528$ which agrees with the experimental ratio of 1,558 with a ~ 2% error.

To explain the cause for the small error between the theoretical and experimental values, and the decay of the W and Z bosons, we consider the modified equation

$$i \partial \Psi(t, \mathbf{r}, R)/\partial t = (\sigma_3 \otimes I_3) \nabla_r \times \Psi(t, \mathbf{r}, R)$$
$$+ \left( \sum_{n=1}^{3} Q_n \sigma_1 \otimes \Lambda_n + (\alpha_0 + i\alpha) \sigma_1 \otimes \Lambda_0 + (\beta_0 + i\beta) \sigma_2 \otimes \Lambda_9 \right) \nabla_R \times \Psi(t, \mathbf{r}, R) \tag{8A}$$

he above equation differs from Eq. (6A) in two additional terms involving $(\alpha_0 + i\alpha) \otimes \Lambda_0$ and $(\beta_0 + i\beta) \sigma_2 \otimes \Lambda_9$, which represents a small perturbation with a non-Hermitian matrix for weak interactions. By taking the 2nd-order time derivative of the modified equation and the Fourier transform one obtains a 6x6 matrix equation om Eq. (8A). By matrix diagonalization and solving the eigenvalue problem, we obtain

$$m_W = (1+\alpha_0)QK\sqrt{1-\alpha^2}(1+(2\alpha/(1-\alpha^2))^2)^{1/4}\cos(\phi_W/2)$$
$$\Gamma_W/m_W = \tan(\phi_W/2), \phi_W = \tan^{-1}(2\alpha/\sqrt{1-\alpha^2}) \qquad (8B)$$
$$m_Z = (2/\sqrt{3})(1+\beta_0)QK\sqrt{1-\beta^2}\left(1+(2\beta/(1-\beta^2))^2\right)^{1/4}\cos(\phi_Z/2)$$
$$\Gamma_Z/m_Z = \tan(\phi_Z/2), \phi_Z = \tan^{-1}(2\beta/\sqrt{1-\beta^2})$$

For a small $\alpha$ and $\beta$ we obtain the ratios for the mass and decay width of the W and Z bosons as

$$m_W/m_Z = \left(\sqrt{3}(1+\alpha_0)/2(1+\beta_0)\right)QK(\cos\varphi_W/\cos\varphi_Z) \approx \sqrt{3}/2$$
$$\Gamma_W/\Gamma_Z = \left(\sqrt{3}(1+\alpha_0)/2(1+\beta_0)\right)(\sin\varphi_W/\sin\varphi_Z) \approx \sqrt{3}/2(\varphi_W/\varphi_Z) \qquad (8C)$$

If $\alpha_0 = \beta_0 = 1, \alpha = \beta \to 0$ for the ideal case $\Gamma_W/\Gamma_Z = \sqrt{3}/2 = 0.866$ is predicted, with no adjustable parameters, as compared to the experimental $(\Gamma_W/\Gamma_Z)_{exp}/2.495 = 0.836 \pm 0.017$.[20] The small errors from the experiments can be accounted for if one includes a perturbation term involving $\alpha$ and $\beta$. Based on the experiments, we obtain $\Gamma_W/m_W \approx \alpha = 0.026$ which is very close to $\Gamma_Z/m_Z \approx \beta = 0.027$. This weak force with a strength ratio of ~ 3% breaks the SU(3) symmetry to cause a small mass shift and decay.

In summary, we propose a chiral preon model to offer physical insights into the electroweak interactions of elementary parties and to derive the mass ratios of the weak and Higgs bosons. Based on Einstein's Pythagorean energy reaction $E = P_1^2 + P_2^2 + P_3^2 + m_0^2$ and Dirac's approach of treating $E$ and $P_{1k}$ as operators in 4D spacetime, we treat mass as an operator ***M*** in the 5th dimension to describe a particle's internal dynamic energy. Using the Gell-Mann SU(3) generators to describe spin-spin exchange couplings to represent the strong nuclear force. Using such a paired chiral preon model, we determine the masses of the photon singlet, Z and W weak boson triplet, and the Higgs boson as composite particles. This analysis indicates that the mass of the weak bosons is acquired from the internal strong spin-exchange couplings between the chiral preon pair. We also show that the symmetry breaking by the weak interaction results in decay and a small shift from the predicted ratios of the mass and decay width. This model differs from the conventional Standard Model based on the Yang-Mils and Higgs mechanisms. With no adjustable parameters, we theoretically derive $m_W/m_Z = \sqrt{3}/2 \sim 0.87$ vs. 0.88, a Weinberg angle of $30^0$ vs. $29^o$, decay width $\Gamma_W/\Gamma_Z = \sqrt{3}/2 \sim 0.87$ vs. $0.84 \pm 0.02$, and a Higgs boson as a composite of W and Z bosons with $m_H/m_W = \sqrt{7/3} \sim 1.53$ vs. 1.56, experimentally.

Our model of a preon pair is based on dualism. Mathematically speaking, a quartet can be constructed from two doubles, leading to a singlet and a triplet as $\mathbf{2 \times 2 = 3 \oplus 1}$, two triplets form an octet and a single as $3 \times 3 = 8 \oplus 1$. Physically speaking in this work, 2 represents a chiral preon pair, 3 represents a triple for the weak boson triplet, and 1 is the photon singlet. For the octet, 8 represents an octet of gluons and 1 as a Higgs boson. The number 8 was used in Gell-Mann's "eight-fold way" quark model. We believe one can use the chiral preon pair to construct the fundamental building block for all composite particles. From a boson-type preon pair with real-valued wave functions, a spinor preon can be constructed. Then layer by layer, a quaternion can be built from paired spinor-type preons, an octonion from paired quaternions, a sedenion from paired octonions. Our view is also supported by the mass ratio between the top quark and the Higgs boson which follows an empirical formula we obtained as $m_T/m_H = 4\sqrt{3}/5 \pm 0.01\%$. Unlike Yang-Mills theory, which assumes point-like particles, our paired rpreon model incorporates a curl operator and Gell-Mann's matrices. To describe a fermion-type preon we follow Dirac's approach in using four anti-commutative operators, our model involves generators of $U(1) \otimes SU(2) \otimes SU(3)$, which is a subgroup of SU(5) proposed by Georgi-Glashow[21], and SO(10) is y Gieorgi[22] for the grand unification theory, and is related to the sedenion algebra model we reported[23]. In contrast, octonion algebra contains SU(3). In this work, we show how a massless particle by coupling to the sedenion operators could acquire an effective mass from the kinetic energy in its internal dynamics. Together with this work, our mass-acquiring mechanism has answered the mass-gap problem.[24] The strong nuclear force reflects the spin-spin coupling because the paired pspin-1/2 preons of opposite chirality behave like two magnetic dipoles. The dipolar interaction energy has $1/r^3$ dependence, yet the Coulomb interaction has $1/r$ dependence. Using an upper limit of quark's size of $\sim 10^{-15}$ m, for two electrons with $10^{-15}$ m separation, the energy due to magnetic dipolar interaction is ~ 36 GeV vs. 1.4 MeV due to Coulomb interaction. This magnetic dipolar energy can reach TeV at ~ 0.4 fm. Unlike dyons with a magnetic charge, preons are not magnetic monopoles, the magnetic dipolar interaction between the paired preons could be the origin of the strong interaction. Extension of this work to other Standard Model particles and their spacetime topology, and hyper-complex algebra deserve further investigation.